\documentclass[aps,prl,twocolumn,showpacs]{revtex4}
\usepackage{amssymb}
\usepackage{amsfonts}
\usepackage{amsmath}
\usepackage{graphicx}
\begin{document}


\title{Gap Anisotropy in Iron-Based Superconductors: A Point-Contact Andreev Reflection Study of BaFe$_{2-x}$Ni$_{x}$As$_2$ Single Crystals}

\author{Cong Ren$^{1,\dag}$, Zhao-Sheng Wang$^1$, Zhen-Yu Wang$^1$, Hui-Qian Luo$^1$,
Xing-Ye Lu$^1$, Bin Sheng$^1$, Chun-Hong Li$^1$, Lei Shan$^1$, Huan Yang$^{1,2}$, and Hai-Hu Wen$^{1, 2}$}

\affiliation{$^1$ National Laboratory for Superconductivity, Institute of Physics and Beijing National
Laboratory for Condensed Matter Physics, Chinese Academy of Sciences, P.O. Box 603, Beijing 100190, China}

\affiliation{$^2$ Physics Department, Nanjing University, Nanjing 210093, Jiangsu, China}

\begin{abstract}
We report a systematic investigation on $c$-axis point-contact
Andreev reflection (PCAR) in BaFe$_{2-x}$Ni$_x$As$_2$
superconducting single crystals from underdoped to overdoped regions
(0.075 $\leq x\leq 0.15$). At optimal doping ($x=0.1$) the PCAR
spectrum feature the structures of two superconducting gap and
electron-boson coupling mode.  In the $s\pm$ scenario, quantitative
analysis using a generalized Blonder-Tinkham-Klapwijk (BTK)
formalism with two gaps: one isotropic and another angle dependent,
suggest a nodeless state in strong-coupling limit with gap minima on
the Fermi surfaces.  Upon crossing above the optimal doping ($x >
0.1$), the PCAR spectrum show an in-gap sharp narrow peak at low
bias, in contrast to the case of underdoped samples ($x < 0.1$),
signaling the onset of deepened gap minima or nodes in the
superconducting gap. This result provides evidence of
the modulation of the gap amplitude with doping
concentration, consistent with the calculations for the orbital
dependent pair interaction mediated by the antiferromagnetic spin fluctuations.
\end{abstract}


\maketitle

\newpage

It is generally accepted that superconductivity in iron pnictides results from a superexchange repulsion mediated by magnetic excitations, which couple electron and hole pockets of the Fermi surface \cite{MazinPRL,LeeDH,Kuroki}.  Such pairing interactions favor either isotropic $s$-wave order parameters with opposite signs on different sheets of the Fermi surface (FS)($s\pm$ model) or anisotropic $s$-wave or even $d$-wave order parameters with nodes \cite{Kuroki2,Graser}.   Consensuses have been reached on several systems, e. g.
LaFePO \cite{Flet}, KFe$_2$As$_2$ \cite{LiSY}, BaFe$_2$(As$_{1-x}$P$_x$)$_2$ \cite{Hashi}, and so on,
that nodes exist on the gap structure.  However,
experimental confirmations of such a nodal-gap state remains highly
controversial in other systems
\cite{DingH,ARPES,ZhangXH,Hardy,Taillefer,Prozorov1,Prozorov2,Imai,MuG,Fukazawa,Raman,WuD}.
For example, measurements of the electronic specific heat of
Ba(Fe$_{1-x}$Co$_x$)$_2$As$_2$ have shown a field dependence
consistent with both a fully gapped FS \cite{Hardy} and a
nodal quasiparticals at the Fermi level \cite{Imai,MuG}.  Such
scattered experimental results and interpretations may come from the
different qualities and doping level of the samples studied.

\begin{figure}
\includegraphics[scale=0.20]{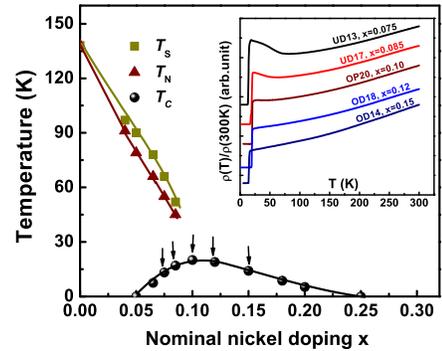}
\caption{\label{fig:fig1}(Color online) Phase diagram of
BaFe$_{2-x}$Ni$_x$As$_2$ as  a function of Ni concentration $x$. The
orthorhombic phase below $T_s$ and the antiferromagnetic (AF) phase
below $T_N$ are also shown here.  The arrows indicate the doping
levels of the samples under investigation.  Inset: Temperature
dependence of the in-plane resistivity $\rho$ for samples with the
Ni nominal doping level $x$ as labeled. Data are vertically shifted
for clarity.}
\end{figure}

Point-contact Andreev reflection (PCAR) spectroscopy has been
adopted for probing  the density of state (DOS) of superconductors
with the high energy resolution. In addition, the capability of this
technique to study the anisotropy and the temperature dependence of
the superconducting gap make it a unique tool in providing
invaluable information for various mechanisms of unconventional
superconductivity (for a review, see Refs. 21, 22). Several theoretical
calculations have been reported on the PCAR conductance
characteristics of a junction involving the $s_{\pm}$ symmetry in
iron pnictide superconductors \cite{Golubov, WangQH}. However, due
to the long-standing issue of surface or/and interface degradation,
experimental results by PCAR technique reveal a wide variation in
the measured Andreev conductance spectra and consequently, the gap
values, especially for the case of $c$-axis junctions
\cite{Sameuley,LGreene}. In this Letter, we fabricate highly
transparent $c$-axis direct contacts to perform the PCAR
spectroscopy study on a series of electron-doped
BaFe$_{2-x}$Ni$_x$As$_2$ single crystals over a wide doping range.
The conductance spectra show a systematic and consistent behavior
with the variation of the doping level, indicative of a doping
dependence of the order parameter for Ni-122 superconductor. In the
$s_{\pm}$ scenario, by using a generalized two-gap
Blonder-Tinkham-Klapwijk (BTK) model, we estimate the gap amplitude
on the hole and electron FS sheets.

High-quality single crystals of
BaFe$_{2-x}$Ni$_{x}$As$_2$ were grown from a self-flux method, as
described elsewhere \cite{Luo}. The crystals were characterized using
x-ray diffraction and energy dispersion (EDX). The doping level in
the crystals was determined by inductive coupled plasma emission
spectrometer (ICP), which gave a Ni concentration roughly 0.8 times
the nominal content $x$.  We choose five compositions: underdoped,
$x$=0.075 (UD13), 0.085 (UD17); overdoped, with $x$=0.12 (OD18),
0.15 (OD14), and optimally doped with $x=0.1$ (OP20). The typical level of impurity phases has been checked
by specific heat measurement on the optimally doped crystal $x=0.1$,
in which a residual component $\gamma_0$ at $T\rightarrow 0$
revealed an impurity phases of $\sim 4\%$ \cite{Wen-Ni}. The
temperature dependence of resistivity for these five compositions
under investigation is displayed in inset of
Fig. 1, by which the bulk transition temperature $T_c$ is determined
(95\% of the normal state resistivity) for each composition.
Consequently, the $T_c$ value for each composition is shown on the
phase diagram in the main panel of Fig. 1.

\begin{figure}
\includegraphics[scale=0.25]{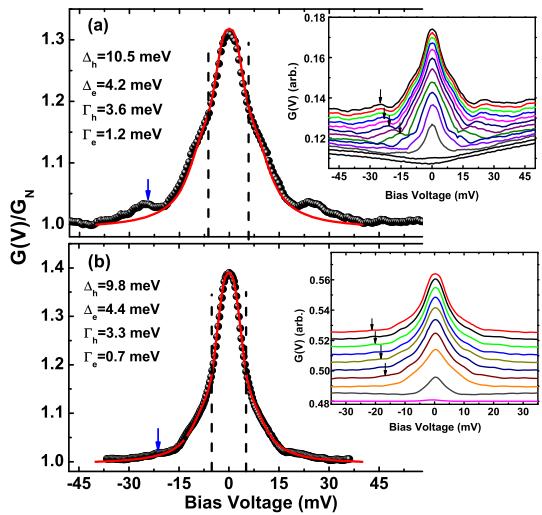}
\caption{\label{fig:fig2} Normalized conductance curves at $T=2$ K
for $c$-axis  contacts (a) OP20a and (b) OP20b.  The red solid lines are
their two-gap fits with the relevant fitting parameters. The blue arrows indicate the additional conductance peak at the edge of gap and the dashed lines mark the ``kink'' structures in the main-gap $G(V)$ curves.
Insets: the raw $G(V)$ curves in temperatures from 2 K to 21 K in a step of 2 K. Data are vertically shifted for clarity. The black arrows mark the corresponding edge-gap conductance peaks in the raw $G(V)$ curves for OP20a and OP20b, respectively.}
\end{figure}

Point contacts to the flat and shiny surfaces cleaved along the
$c$-axis of  BaFe$_{2-x}$Ni$_x$As$_2$ crystals were made using thick
silver paste (4929N DuPont) bonding with gold wires (of 16 $\mu$m
diameter). The typical size of these planar contact is about
0.08-0.15 mm under a microscope.  Due to the nanocrystalline nature
of the silver paint, the contact made in this way, is actually
formed by many nanocontacts analog to tip point-contact technique
\cite{Italy}.  For the backside electrical wiring,
we applied ultrapure indium or silver paste to cover the whole area
of the bottom surfaces of the crystals. On each
piece of the crystal, 5-6 planar contacts were made from point to
point to ensure the reproducibility and consistency of the junction
conductance spectra and their spectroscopic nature.

Fig. 2(a) and (b)  show the raw (inset) and normalized conductance curves $G(V)=dI(V)/dV$
of two $c$-axis Ag/BaFe$_{1.9}$Ni$_{0.1}$As$_2$ point contacts (OP20a and OP20b),
respectively.  The contacts made in this way remain stable in
thermal cycling, and the contact resistance at high bias $R_N$
varies very little ($< 6$\%) over the whole $T$ range up to $T_c$.
The Andreev signal as the conductance enhancement decreases on
increasing $T$ and vanishes at $T\geq T_c$, leaving a slightly
asymmetrical $V$-shaped normal state.  Shown in the main panels of
Fig. 2(a) and (b), the magnitude of the Andreev reflection reaches
as high as 30\%-40\%, implying a relatively transparent boundary
between Ag nanoparticle and BaFe$_{2-x}$Ni$_{x}$As$_2$
superconductors. The stabilities in $R_N$ and the high level of Andreev signal indicate that the
conduction channels through the contact is in ballistic regime, and
therefore, energy-resolved spectroscopy is possible.  A feature shows up in these conductance curves: an additional peak at $\sim 20 $ mV, and the peak gradually disappears with $T$ approaching $T_c$.
It seems that this peak is much pronounced when the Andreev signal is relatively low, which is close to the case of
tunneling side (see below).  Very recently, this conductance peak at the edge of the gap has been observed in Co-122 crystals \cite{Italy2}, and is attributed
to the signature of an electron-boson coupling associated with the superconducting gap.  The observation of the electron-boson coupling mode in the conductance spectra implies the high quality of the point-contacts and thus their spectroscopic nature.

The two-gap superconductivity manifests itself as a ``kink'' in the in-gap conductance, marked at the dashed lines in Fig. 2 (a) and
(b).  To explicitly describe the variety of spectral behavior observed and quantitatively resolve the gap amplitude, we invoke a
generalized BTK formula \cite{BTK} with three parameters: a dimensionless parameter $Z$ which represents the interface
transparency; an imaginary quasiparticle energy modification $\Gamma$ \cite{Dynes} which reflects the spectral broadening, and
the superconducting gap $\Delta$. In BTK model, the normal and Andreev reflection probabilities, respectively, are related to the DOS of the superconductor $N_s=N_0 \textrm{Re}(\frac{E-i\Gamma}{\sqrt{(E-i\Gamma)^2-\Delta^2(T,\theta)}})$ with $N_0$ the normal-state DOS and $\theta$ the crystalline angle parallel to the current injection.  To choose a gap function to calculate these two-gap conductance spectra, we assume, based on the $s_{\pm}$ scenario, an isotropic gap $\Delta_h$ and an anisotropic gap of the general form $\Delta_e[1-r+r\cos(2\theta)]$, with the gap anisotropy ratio $r$ varying from $r=0$ (isotropic $s_{\pm}$ state) to $r=1$ (completely $d$-wave) \cite{Chubukov,MazinRaman,WuD}. Therefore, by the standard two component conductance (current) model the conductance spectra is the contributions of the hole-like ($G_{h}$) and electron-like ($G_e$) Fermi pockets: $G = wG_{h} + (1-w)G_{e}$, where $w$ is the spectral weight. For simplicity, we assume a balanced contribution of hole and electron Fermi surfaces to the spectral conductance by taking $w$=0.5.

\begin{figure}
\includegraphics[scale=0.28]{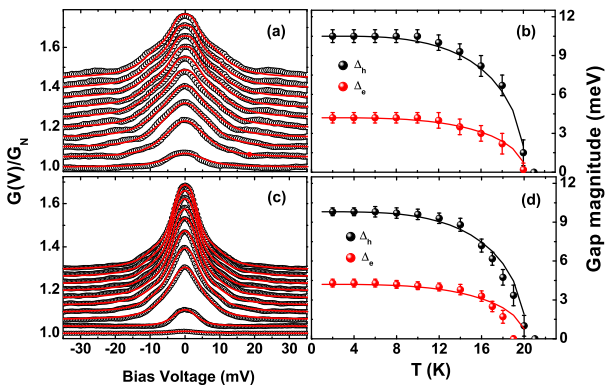}
\caption{\label{fig:fig3}(Color online) Temperature dependence of the normalized conductance
spectra and the relevant two-gap BTK fits (solid red lines) for (a) OP20a, and (b) OP20b, respectively.  Data and their fits are vertically shifted for clarity except the bottom ones.  The step
of $T$ increase is 2 K.  The obtained gap magnitude for (b) OP20a, and (d) OP20b, respectively, as a function of $T$.   The solid lines are the fits to an empirical gap function, see text. }
\end{figure}

Examples of normalized $G(V)$ curves and their fits at $T=2$ K are shown in the main panels of Fig. 2(a) and (b) for junctions OP20a and OP20b,
respectively. The two-gap BTK model (red lines) fit very well the main features of the experimental $G(V)$ curves except the
electron-boson coupling mode around 20 meV, yielding a set of fitting parameters associated with gap magnitude and anisotropy ratio: $\Delta_h=10.5$ meV, $\Delta_e$=4.2, and $r$=0.3 for OP20a and $\Delta_h=9.5$ meV, $\Delta_e=4.5$ meV, and $r$=0.3 for OP20b, respectively. It is noted that the same gap magnitude is also extracted from a recent PCAR experiment on a $c$-axis Ag/BaFe$_{1.8}$Co$_{0.2}$As$_2$ with the comparable $T_c=24$ K \cite{Italy2}.  Here we emphasize that a two-gap formula with two $s$-wave gap ($r=0$) can also fits our experimental data rather well. However, the parameters $\Gamma/\Delta=0.6-0.7$ are applied to fulfill the fit at this low $T$, which brings a large uncertainty in the gap magnitude.

With these fitting parameters, we check the validity of these fits by extending the fit to the overall temperature spectral. As shown in Fig. 3(a) and (c), the two-gap $s_{\pm}$ model still fits reasonably well the $T$-dependence of these $G(V)$ curves with fitted gap magnitude.  In this overall-$T$ spectral fit, $r=0.3$ and $Z_{h(e)}=0.3-0.2$ are constant with $T$ while $\Gamma_h=3.3$ and $\Gamma_e=1.3$ meV (OP20a) and $\Gamma_h=2.9$ and $\Gamma_e=0.8$ meV (OP20b) are almost constant or slightly increase with $T$. From the fits of various curves we obtain the gaps $\Delta_h$ and $\Delta_e$ as a function of $T$, which is plotted in Fig. 3(b) and (d) for these two junctions, respectively.  For comparison, the obtained gaps can be approximated by an empirical gap formula: $\Delta(T)=\Delta_0\tanh(\alpha\sqrt{T_c/T-1})$ with $\alpha=1.95$ for $\Delta_h$ and 1.86 for $\Delta_e$ (cf. $\alpha=1.74$ for weak-coupling BCS gap).

We analyze the physical meanings of the obtained gap values and gap function.  It is shown from angle-resolved photoemission spectroscopy experiment on a Co-122 crystal that the large gap $\Delta_h$ is located on the hole FS sheet, instead, the small gap $\Delta_e$ is presented on one of the electron FS sheets \cite{ARPES}.  The gap values $2\Delta_h/k_BT_c\approx 11.6$ and $2\Delta_e/k_BT_c\approx 5.0$, both above the BCS weak-coupling ratio. Besides, the $\alpha$ value from the $\Delta(T)$ function also points to a strong-coupling character for both $\Delta_h$ (hole FS) and $\Delta_e$ (outer electron FS).  These results are consistent with a three-band $s_{\pm}$ Eliashberg model \cite{threeTHY}, in which spin fluctuations mainly provide the interband coupling, and thus so in the electron-boson coupling matrix.  On the other aspect, the existence of strong electron-boson coupling in this compound is manifested by the observation of the spectral peak $E_p$ at about 20 mV. In our low-transparency (large $Z=0.3$ for OP20a) point contact, a characteristic energy of $\Omega_b=E_p-\Delta_{max}=13$ meV and 11 meV (OP20b junction with small $Z=0.2$).  This energy scale is compatible with the spin-resonance energy observed by neutron scattering on the same crystals \cite{a1}.

The obtained anisotropy ratio $r=0.3$, resolved in our $c$-axis PCAR spectroscopy of BaFe$_{1.9}$Ni$_{0.1}$As$_2$, indicates a full gap state with gap minima along $c$ axis.  This nodeless state of optimally doped Ni-122 is in similarity with that of optimally-doped Co-122, in which a gap minima is already present at maximal $T_c$ by the $c$-axis thermal conductivity measurements \cite{Taillefer,Taillef}.

It is natural, however, to inspect the manner of the superconducting gaps in the crystals with doping away from the optimum.  We have measured the point-contact $G(V)$ curves in whole $T$ range up to $T_c$ for junctions with $x=0.075$ (UD13), 0.085 (UD17), 0.12 (OD18), and 0.15 (OD14).  The typical $G(V)$ curves at $T=2$ K ($<0.2\ T_c$) and $T\geq T_c$ are shown in Fig. 2(a)-(d) for these four samples respectively.  As shown, these $G(V)$ curves exhibit a consistent behavior: 1) An underlying feature of a \emph{dominant} single gap is unambiguously identified with a similar conductance enhancement of 25\%-35\% for each
junction; 2) A parabolic normal-state $G(V)$ curve with a slight asymmetry at $T\geq T_c$ for each $x$, opposite to those of hole-doped K-122
\cite{LGreene,ZhangXH}, implies the similar origin of the underlying normal-state background.  Nevertheless, a striking feature in these normalized (and the raw) $G(V)$ curves is that: at $T=2$ K, a conductance plateau and/or a double peak around zero bias for junctions UD13 and UD17 gradually evolutes into an in-gap sharp peak in $G(V)$ for junctions OD18 and OD14. Considering the overall spectral consistency in these junctions, the systematic evolution of the Andreev conductance
spectra with doping concentration is nontrivial. Qualitatively, \emph{for highly transparent junctions at finite $T$}, the appearance of an in-gap
plateau in Andreev conductance spectrum is a signature of a fully gapped state. In contrast, an in-gap conductance peak is a
characteristic of an anisotropic gap state due to \emph{the presence of a finite DOS at low energy}, like a $d$-wave gap in cuprates
\cite{Deutscher}.  We note that it is not easy to describe the spectral behavior using simple formulism, because we are dealing with multi-band or/and even multi-gap system. Therefore, our observation that the systematic evolution from the in-gap conductance plateau for the underdoped samples to the in-gap peak in $G(V)$ curves for the overdoped samples indicates the existence of doping induced
evolution of superconducting gaps with an isotropic feature in the underdoped region to an anisotropic, even, nodal gap in the
overdoped side. This is highly consistent with the result of the $T$-dependent penetration depth $\lambda$ in a series of Ni-122
superconductors, in which $\Delta\lambda \propto T^n$ with the exponent $n\geq 2$ for underdoped samples and $\Delta\lambda$
becomes more linear-$T$ dependent for overdoped samples, indicating the development of nodal gaps in the overdoped region
\cite{Prozorov1}.

\begin{figure}
\includegraphics[scale=0.3]{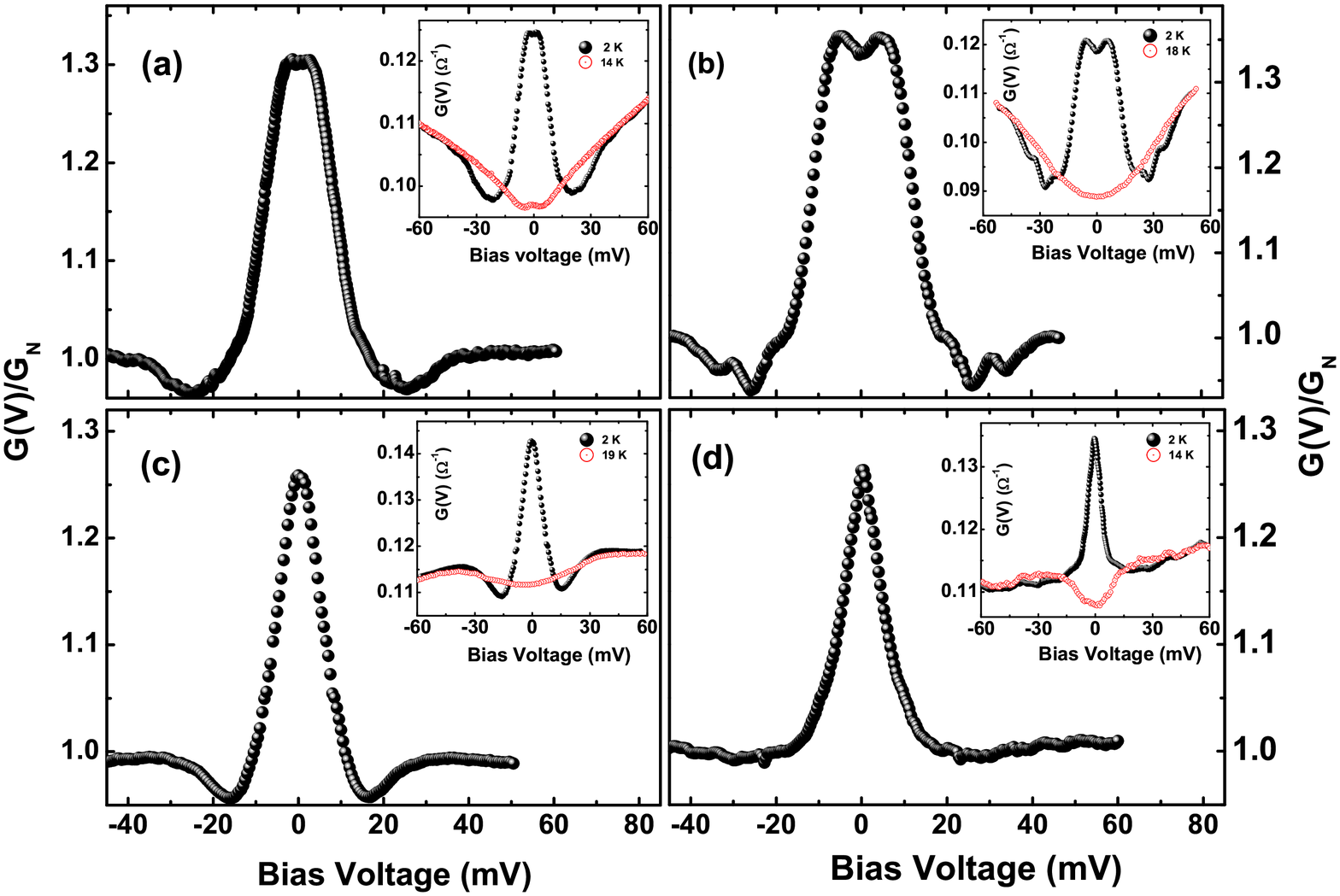}
\caption{\label{fig:fig4}Normalized conductance curves at $T=2$K for (a) UD13, (b) UD17, (c) OD18, and (d) OD14.  Insets: The corresponding raw data of conductance spectra at 2 K and $T\geq T_c$. }
\end{figure}

In conclusion, measurements on point-contact junctions made on
single crystals of BaFe$_x$Ni$_{1-x}$As$_2$ illustrate an
interesting evolution of the gap structure. The Andreev conductance
spectra clearly show a full-gap state for underdoped crystals and a
highly anisotropic, perhaps nodal-like gap state for overdoped
crystals. Quantitative analysis of the spectral data of
optimally-doped contacts using a generalized BTK formalism resolves
two superconducting gaps in strong coupling limit. Resulted from the
analytical fitting, the small gap on the electron-like FS sheets
shows a crossover from a nodeless in the underdoped side to a nodal
feature in the overdoped region.  This result provides evidence of
the modulation of the gap amplitude on the FS with doping
concentration, consistent with the calculation for the orbital
dependent pair interaction mediated by the antiferromagnetic spin fluctuations.

Acknowledgement: The authors are grateful to Profs. R. Prozorov,
Qiang-Hua Wang and Dr. Gang Mu for intensive discussions. This work
is supported by the National Science Foundation of China, the
Ministry of Science and Technology of China (973 project No:
2011CBA00100), and Chinese Academy of Sciences (Project ITSNEM).

$\dag$cong\_ren@iphy.ac.cn

\end{document}